\documentstyle{article} 
\begin{document} 
\begin{flushright} LPTHE Orsay-98/53
\end{flushright}
\begin{center}
{\Large Is the resonance D(2637) really a radial excitation?} 
\\
D. Melikhov
\footnote{On leave of absence from Nuclear
Physics Institute, Moscow State University, Moscow, 119899, Russia} 
and O. P\'ene
\\
Laboratoire de Physique Th\'eorique et Hautes Energies, \\
Universit\'e de Paris XI, B\^atiment 211, 91405 Orsay Cedex, France
\footnote{
Laboratoire associ\'e au Centre National de la Recherche Scientifique - URA D00063\\
{e-mail: melikhov@qcd.th.u-psud.fr, pene@qcd.th.u-psud.fr}}
\end{center}
\begin{abstract} 
We consider various possible identifications 
of the quantum numbers of the resonance $D(2637)$ 
recently observed by DELPHI in the $D^*\pi\pi$ channel. 
We argue that in spite of a good agreement of the measured mass with the 
quark-model prediction for the radial excitation, a  total width 
as small as $\le 15$ MeV is hardly compatible with its identification as a 
radial charm excitation. 
The $J^{P}=2^{-}, 3^{-}$ orbitally excited mesons with such a mass 
could have  widths of the observed order of magnitude. However in this case 
one would expect two neighbouring states with the mass
difference of about 30-50 MeV corresponding to the nearly degenerate components 
of the heavy-meson multiplet
with light-quanta angular momentum j=5/2 , and moreover, according to the quark-model 
predictions the mass of the orbital excitation should be more than 50 MeV larger than 2637
MeV. 
Thus we conclude that, at present, we find no fully convincing understanding of the 
quantum numbers of the observed resonance. 
\end{abstract}
\vskip1cm
Recently, DELPHI has observed a narrow resonance $D(2637)$ in the $D^*\pi\pi$
channel with a total width of less than the detector resolution: $15$ MeV \cite{delphi}.
 The mass of the observed resonance
turns out to be in perfect agreement with predictions of the quark models 
\cite{gi,efg} for the charm radial excitation $D^*{'}(J^P=1^-)$. 
This coincidence has lead to a quick identification of the discovered resonance state 
with the radially excited vector charm meson. 

In this letter we reconsider the identification of the quantum numbers of the observed 
resonance by submitting it to the following criteria: 
\begin{itemize}
\item[i] the resonance mass should be $\simeq$ 2637$\pm$6 MeV; 
\item[ii] the resonance width should be $\le 15$ MeV; 
\item[iii] the resonance should have a sizeable branching ratio of the channel $D(2637)\to D^*\pi\pi$ 
in which  it has been observed. 
\end{itemize}

We estimate the decay rate of a radial excitation of the reported mass and 
find that, 
although the partial widths are strongly model-dependent, the total rate is 
conservatively estimated to be significantly larger than about 50 MeV. 
We thus conclude that the observed width of only 
$\le 15$ MeV is hardly compatible with its identification as a radial charm 
excitation. 

An identification of the observed resonance with an orbital charm excitation 
seems to be 
more favourable: two mesons with the quantum numbers $J^P=2^-,3^-$ 
(quark orbital momentum
$L=2$) could have a width 
of the observed order of magnitude. However,  theoretical estimates yield a mass 
for the latter orbitally-excited states approximately $\ge 50$ MeV above 
the reported value. 
In addition, in this case, one would expect {\it }two neighbouring states with a mass 
difference of about 30-50 MeV (similar to the mass difference between 
$D_{2^+}(2460)$ and $D_{1^+}(2420)$ in the j=3/2 positive-parity sector), 
while only 
one resonance has been reported. Altogether we conclude that,
 at the moment, there is no fully convincing understanding of the 
quantum numbers of the observed resonance. 

Our analysis is based on combining the heavy-quark symmetry relations for the 
transition amplitudes between heavy mesons through the emission of the light 
hadrons with the quark-model estimates. Namely, 
we estimate the decay rates of radially and orbitally 
excited charm resonances into $D\pi$, $D^*\pi$, $D\pi\pi$ and $D^*\pi\pi$ 
assuming the resonance mass of 2637 MeV as measured by DELPHI. 

To obtain estimates of the branching ratios of 
the three-body decays with two pions in the final state we treat them as cascade 
two-body decays $D(2637)\to (D,D^*)R\to (D,D^*)\pi\pi$
through the intermediate Breit-Wigner resonance with relevant quantum numbers. 
The status of the $(\pi\pi)_{l=0, I=0}$ channel is not well-defined, and we varied the 
corresponding $\sigma$ resonance mass in the range $400\div800$ MeV and the width 
in the range $\Gamma(\sigma)\simeq 700\div900$ MeV. The low-energy $(\pi\pi)_{l=1}$ 
partial-wave is dominated by $\rho$. Higher partial waves of the $\pi\pi$ system give 
negligible contributions. 

Table \ref{table:1} lists the candidate 
charm states and their allowed decay modes as given by the spin-parity conservation. 

For  heavy-meson decays, additional constraints are given  by the heavy-quark (HQ) 
symmetry \cite{iw}. Namely, in the heavy quark limit the heavy quark spin decouples from other
degrees of freedom and remains conserved in hadron transitions. Thus, in strong decays of
heavy hadrons, the total angular momentum of the heavy and light degrees of freedom are
conserved 
separately, in addition to the conservation of the parity and total 
angular momentum. Hence, with respect to strong decays, heavy hadrons can be
assigned an additional conserved quantum number, $j$, which is the total angular momentum of the
light degrees of freedom. The consequences of the HQ symmetry for hadron transitions have been worked
out by Isgur and Wise \cite{iwprl}. Namely, the HQ symmetry allows one to relate to each other 
different amplitudes of strong transitions between the states with fixed $j$ and $j'$, the
latter being the angular momenta of the light degrees of freedom in the initial and final
hadronic states, respectively. Table \ref{table:1} also presents the HQ symmetry allowed transitions
in terms of the few independent amplitudes. The $O(1/m_Q)$ corrections in the effective Hamiltonian  
yield corrections to these HQ symmetry relations. However, for our order-of-magnitude 
analysis these corrections are generally unimportant and will ne neglected unless explicitly
specified. 

For the calculation of the independent amplitudes ($\alpha$, $\beta$, $\delta$, and $\xi$ in Table 
\ref{table:1}) one needs  a non-perturbative approach. 
We apply here a naive quark-pair-creation (${}^3P_0$) model \cite{lopr} which, in spite
of its simplicity, has proven to provide a reasonable quantitative description of the two-body hadronic 
decays. The model is based on the assumption that the the spectator quarks do not change their 
SU(3) quantum numbers, nor their momenta and spins. The created quark-antiquark pair should be
therefore in a ${}^3P_0$ ($J^{PC}=0^{++}$) SU(3) singlet state of zero total 3-momentum. More details 
concerning the model can be found e.g. in Ref. \cite{lopr}. 
To compute the transition amplitudes one needs an overall transition-strength constant 
which determines the amplitude of the production of the light $q\bar q$ pair from the vacuum and 
the wave functions of the initial and final mesons. 
For the overall strength constant we use the value $\gamma=2.2$ as found from the analysis of 
the hadronic decays in the light sector \cite{lopr}. 
For the nonperturbative meson wave functions we assume an harmonic-oscillator approximation. 
The remaining parameters to be fixed are  the size of the light-light and heavy-light wave
functions. They are given by two radii 
the light-light ($R$) and heavy-light ($R_D$). For the ground state 
these radii are simply $R^2 =2/3 <(\vec r_q -\vec r_{\bar q})^2 >$. 

The heavy meson has to be smaller than the light one. From estimates with different
potentials, we assume
 the radii to satisfy the relation $R^2_D/R^2\simeq 0.5\div0.7$ and 
allowed $R^2=6\div9$ GeV${}^2$: such values of $R$ are compatible 
with previous descriptions of the spectrum and  decay rates  \cite{lopr} 
and in addition, we  check, Table \ref{table:3},  that the 
${}^3P_0$ model with these parameters describes correctly
 the experimentally observed 
$D_{2^+}\to (D,D^*)\pi$ decay rates.  
For the $D^*\to D\pi$ transition, Table \ref{table:2}, we also express the decay width in
terms of a dimensionless coupling constant defined as follows
\begin{equation}
\label{decconst} 
\langle D^0(p_2)\pi^+(q)|D^{*+}(p_1)\rangle =g_{D^*D\pi}q_\mu\epsilon^\mu_1, 
\end{equation}
where $\epsilon^\mu(p_1)$ is the vector-meson polarization vector, the states
being normalized covariantly and where we have omitted the momentum
conservation delta function. 
The $g_{D^*D\pi}$ of the ${}^3P_0$ model agrees with the 
experimental bound although seems to be a bit small compared with other 
theoretical estimates. Notice however that  
the ${}^3P_0$ model is 
non-relativistic and as such does not describe properly the soft pion limit. In the 
$D^*\to D\pi$ decay the pion is produced almost at rest 
and the model is indeed expected to underestimate the coupling constant as observed. 
But the model has proven to work rather well 
for hadronic resonance decays in the usual domain of the emitted pion energies, i.e.  
$E_\pi\simeq$ 300-600 MeV. Indeed, 
as we will see later,  the model estimates for the decay of a heavier 
radially-excited $D'$ are in better agreement with covariant methods. 

Having thus fixed  the ranges of the basic parameters from the light sector and 
$D_{2^+}\to (D,D^*)\pi$ decays, we apply the model to the analysis of the decay of 
radially and orbitally excited negative-parity states. 

{\bf Radial excitation \boldmath $D_{1^-}$ \unboldmath}

The decay of the vector radially-excited $D_{1^-}$ into $D\pi$ and $D^*\pi$ 
is governed by the following amplitudes 
\begin{eqnarray}
\label{decconst1} 
\langle D^0(p_2)\pi^+(q)|D^{*{}'+}(p_1)\rangle&=&g_{D^*{}'D\pi}q_\mu\epsilon^\mu_1, 
\\
\langle D^{*0}(p_2)\pi^+(q)|D^{*{}'+}(p_1)\rangle&=&
ig_{D^*{}'D^*\pi}\epsilon_{\mu\nu\alpha\beta}p_1^\mu p_2^\nu\epsilon^\alpha_1
\epsilon^\beta_2\nonumber
\end{eqnarray}
with the coupling constants $g_{D^*{}'D\pi}$ and $g_{D^*{}'D^*\pi}$ to be determined on the 
basis of a dynamical approach. 
In the heavy-quark limit the constants are related to each
other as follows 
\begin{eqnarray}
\label{hqsr}
g_{D^*{}'D\pi}=M_{D^*{}'}\;g_{D^*{}'D^*\pi}. 
\end{eqnarray}

To estimate these coupling constants one can applied various theoretical approaches. 
For instance, one may use the PCAC definition of the pion field (although the pion 
is not soft at all in this decay) in which case a complicated problem of calculating 
the coupling constants of interest is reduced to a relatively  simpler one of 
calculating the $D^*{}'\to (D,D^*)$ transition form factors through the axial-vector 
current. Namely, one finds 
\begin{eqnarray}
\label{pcac}
g_{VP\pi}=\frac{1}{f_{\pi}}[(M_V+M_P)A_1(0)+(M_V-M_P)A_2(0)]. 
\end{eqnarray}
For estimating the meson transition form factors $f$ and $a_+$ 
we used the relativistic dispersion approach of 
Ref. \cite{mel} adopting wave functions 
of the ground-state and radially-excited $D$ which provide the values 
$f_D\simeq 200$ MeV, $f_D^*\simeq 240$ MeV and $f_{D^*{}'}\le 400$ MeV. 

Actually, one observes a suppression of the form factor  
$A_1$ in the $D^*{}'\to D$ transition, as compared with the $D^*\to D$ one, 
because of the orthogonality of the wave functions of the orbitally-excited
and the ground states
\footnote{We are indebted to Damir Becirevic and Alain Le
yaouanc for attracting our attention on this suppression which they find even 
stronger in a related approach \cite{damiralain} based on the Dirac equation.}
: namely, $A_1^{D^*\to D}(0)\simeq 0.5$  and 
$A_1^{D^*{}'\to D}(0) \simeq 0.2-0.3$. In the absence of the second term
of (\ref{pcac}) the $D^*{}'\to (D,D^*)$ transition would thus be suppressed by
a factor 2 to 4 in rate. However such a suppression due to orthogonality in the
soft pion limit is expected to be reduced when the momentum recoil (which is
large in this case) is taken into account. The transition $D^*{}'\to (D,D^*)
\pi$ would only produce a soft pion if the mass of the $D^*{}'$ would be close
to that of the $D^{(\ast)}$, thus canceling the second term in (\ref{pcac}). 
But we are not in such a situation, the second term is not negligible since
$A_2^{D^*{}'\to D}(0)\simeq 0.9-2.5$ and one ends up with the relation 
$g_{D^*{'}D\pi}=(0.5-1.5)g_{D^*D\pi}$ where $g_{D^*D\pi}\simeq 15$. 
Notice that the large uncertainties in $g_{D^*{'}D\pi}$ are connected 
with a strong sensitivity of the latter to the subtle details of the 
sign-changing wave function of the radially-excited state. 
Altogether, and using the HQ symmetry relation (\ref{hqsr}) 
we find  a rather large range for our estimate: 
$\Gamma(D^*{}'\to D\pi)=20-200$ MeV and $\Gamma(D^*{}'\to D^\ast\pi)=25-250$ MeV.  

To obtain more precise estimates we  also used the ${}^3P_0$ model and found values in the
region $g_{D^*{}'D\pi}=13\div15$ which are compatible with PCAC based estimates. 
Table \ref{table:4} presents the ${}^3P_0$ model estimates of the $D'_{1^-}(2637)$ decay rates. 

So for the sum of the decay rates of the channels 
$\Gamma(D^*{}'\to D\pi)+\Gamma(D^*{}'\to D^*\pi)$ one finds rather uncertain estimates 
ranging from 40$-$50 MeV, which is only slightly above the reported width of 
the DELPHI resonance, to several hundreds MeV, which is far above. 

Notice that anyway an important contribution to the total width of the 
radially-excited
$D^*{}'$ is given by the decay channel $D^*{}'\to D^*(\pi\pi)_{l=0}$ allowed by the HQ
symmetry in the S-wave (Table \ref{table:1}). The $^3P_0$ model estimate of its rate is 
120-160 MeV. So, conservatively, one cannot expect the total width of the radially 
excited state $D*'_{1^-}(2637)$ to be less than 50 MeV and presumably it should be 
much broader. This has the double effect of predicting a large branching ratio
for the observed $D^\ast\pi\pi$ channel, but  unhappily also a large total
width.
Thus we conclude that 
{\it identification of the resonance $D(2637)$ as a radial excited 
$J^P=1^-$ charmed meson is hardly compatible with the reported value 
of the resonance total width $\le 15$ MeV}.  

{\bf Orbital excitations \boldmath $D_{2^-,3^-}$ \unboldmath} 

To proceed with these states, we assume the HQ symmetry at the level of the transition 
amplitudes but use the physical masses to compute the relevant phase-space factors. 
Namely, we calculate the transition amplitudes of the
modes $D_{3^-,j=5/2}\to D^{(\ast)}\pi$ and $D_{3^-, j=5/2}\to D^{(\ast)}(\pi\pi)_{l=0,1}$ 
and determine all other related amplitudes through the HQ symmetry relations listed 
in Table \ref{table:1}. 

Notice that the transition of $D_{2^-,j=5/2}$ into the four positive parity states
$D_{2^-,j=5/2}\to D^{**}_{j=1/2,3/2^+}\pi$ $\to D^{(\ast)}\pi\pi$, is allowed,
in the heavy-quark limit, only in the D-wave, since the latter positive parity
states with $j=1/2$ or $j=3/2$ cannot be produced with an S-wave pion 
from a $j=5/2$. A rough estimate convinced us that, due to the small 
final momenta, these decays will be even more suppressed than the other 
$D^{(\ast)} \pi\pi$ channels to be discussed later. 
 
Taking into account the phase-space factors yields the 
results listed in Tables \ref{table:5} and \ref{table:6}. 
Some comments on the presented numbers are in order. It can be seen from Table 
\ref{table:1} that the decay into $D^{(\ast)} (\pi\pi)_{l=1}$ can go through
a P-wave between $\pi\pi$ and the charmed meson. Compared to the F-wave for $D^{(\ast)}\pi$
and the D-wave for $D^{(\ast)} (\pi\pi)_{l=0}$, this decay channel has to be
considered
even though only a small part of the $\rho$-meson 
Breit-Wigner tail is included in the phase space. The $D^{(\ast)} (\pi\pi)_{l=0}$ is
suppressed by one order of magnitude as compared to the 
$D^{(\ast)} (\pi\pi)_{l=1}$ due to the centrifugal barrier suppression.   

The physical $D_{3^-}$ state is dominantly the $D_{3^-,j=5/2}$ one, and one can safely 
estimate $\Gamma(D_{3^-})\simeq \Gamma(D_{3^-,j=5/2})$, since a possible small 
admixture of the narrow $D_{3^-, j=7/2}$ state does not practically change its width. 
For the $D_{2^-}$ state however this is not the case: one might expect a sizable increase 
of the width of the physical $D_{2^-}$ with respect to the 
width of the $D_{2^-,j=5/2}$. For example, even a small admixture 
of the $D_{2^-,j=3/2}$ to the dominant $D_{2^-,j=5/2}$ may increase  
the total width of the physical $D_{2^-}$ meson, since 
one expects $\Gamma(D_{2^-,j=3/2})\gg\Gamma(D_{2^-,j=5/2})$ 
due to the $P$-wave decay mode $D_{2^-,j=3/2}\to D^*\pi$ allowed by the HQ symmetry. 

A similar situation has been observed in the positive-parity $D_{j=3/2}$ multiplet: 
namely, the experimental ratio of the decay rate of $D_{1^+}(2420)$ which is dominantly 
$D_{1^+,j=3/2}$ and the decay rate of $D_{2^+}(2460)$ which is practically 
the rate of a pure $D_{2^+,j=3/2}$ is: 
\begin{equation}
\Gamma(D_{1^+}(2420))/\Gamma(D_{2^+}(2460))=0.71\label{ratio1}
\end{equation}
that is about twice larger than the leading-order HQ symmetry estimate 
\begin{equation}
\Gamma(D_{1^+,j=3/2})/\Gamma(D_{2^+,j=3/2})=0.3.\label{ratio2}
\end{equation} 
The latter value is obtained by assuming the leading-order HQ symmetry relations between the amplitudes 
and taking the physical masses of the corresponding states for the calculation of the relevant
phase-space factors. 

This discrepancy can be solved by invoking the $1/m_c$ corrections e.g. by assuming a small 
admixture of a broad $D_{1^+,j=1/2}$ to a narrow $D_{1^+,j=3/2}$ in the physical 
$D_{1^+}(2420)$ as proposed in Ref. \cite{lwi}. 
Another possibility (see \cite{falk} and refs therein) is to have a rather strong 
increase in the decay rate of a pure $D_{1^+,j=3/2}$ due to the $1/m_c$ corrections to the
effective Hamiltonian  
(A combination of these two variants is of course also possible). 
Anyway the $D_{j=3/2}$ positive-parity sector prompts that the net effect of the $1/m_c$ 
corrections is a doubling of the ratio $\Gamma(D_{J=j-1/2,j})/\Gamma(D_{J=j+1/2,j})$. 

Hence the obtained value of the $D_{2^-,j=5/2}$ width should be considered as a lower 
bound of the physical $D_{2^-}$ width. Still, from the comparison with the $j=3/2$ sector, 
we expect the physical $D_{2^-}$ width not to exceed the HQ estimate of the $D_{2^-,j=5/2}$ 
state by much more than a factor 2.  

Similarly, the subleading $1/m_c$ effects can influence also the rate of the
transition $D_{2^-,j=5/2}\to D^{**}_{j=1/2,3/2^+}\pi$. 
A rough estimate based on the eqs. (\ref{ratio1}), (\ref{ratio2}), where 
an S-wave $O(1/m_c)$ decay leads to an increase of about 10 MeV of the width, 
and considering that the phase space is
smaller in the decay of a $D(2634)$ into positive parity resonances than in the
decay of the latter into the ground state, we are not too worried. Still, a 
closer scrutiny of these $O(1/m_c)$ effects would be welcome.

Finally, we conclude that {\it the orbitally excited $D_{2^-}$ and $D_{3^-}$ charm 
resonances with the mass in the region of $2640$ MeV can have the width 
of the order reported by DELPHI}. 
However, an identification of the resonance $D(2637)$ with an orbital excitation in the charm 
system is not straightforward since the reported mass seems to be significantly
 smaller than  the theoretical expectations. 

For example, the Godfrey-Isgur (GI) model \cite{gi} which describes with a good 
accuracy nearly all 
known mesons predicts the mass of the $D_{3^-}$ to be 2830 MeV. Taking into account that 
for the $D_{2^+}$ state the GI model gives 2500 MeV which is 40 MeV heavier than the observed 
value of 2460 MeV, we can expect for the $D_{3^-}$ state the mass $\le$ 2800 MeV. From
 the typical mass-splitting between the states with neighbouring $j=2$ and $j=3$ in the 
GI model, one could expect the $D_{2^-}$ mass near 2750 MeV.  

The quark-gluon string model \cite{kn} has predicted the masses 
of the charm resonances $M(D_{2^-})=2660\pm 70$ MeV and $M(D_{3^-})=2760\pm 70$ MeV. 
The partial rates of the latter were estimated to be $\Gamma(D_{3^-}\to D\pi)=1.3\div 2.0$ MeV 
and $\Gamma(D_{3^-}\to D^*\pi)=3.5\div 7.0$ MeV yielding the total width of $D_{3^-}$ in the
necessary range. 
On the other hand a large mass splitting between the $2^-$ and $3^-$ charm states signals 
that the $D_{2^-}$ in the quark-gluon string model contains a big admixture of the lighter
state $D_{2^-,j=3/2}$. So, for the mass of the  pure $D_{2^-,j=5/2}$ state one would expect a 
higher value. 

Altogether, from the above theoretical analyses we could expect the mass of the $D_{2^-,j=5/2}$ 
in the region of $2650\div2750$ MeV which is only marginally compatible with 
 the reported resonance mass of 2637 MeV.  

The branching ratios of the $D^\ast\pi\pi$ channels in table 5 and 6 are in the
range of a few percent, which may seem a little small for these channels to have
been observed. Still some non-resonant $\pi\pi$ contributions, which are difficult 
to estimate, will add up.  

Summing up, if the resonance $D(2637)$ with the width of $\le 15$ MeV is 
confirmed by further analyses (at the moment CLEO and OPAL do not see it \cite{exp}) 
then the theoretical understanding of its quantum numbers is not clear: 
in spite of the coincidence of the observed mass with the predicted mass of the radial charm 
excitation $D'_{1^-}$, its interpretation as a radial excitation is completely ruled out by 
the small observed width. 

On the other hand, although an identification of the state with the $D_{2^-}$ orbital excitation
seems to be appropriate from the viewpoint of the total width, its mass seems to be too low
compared with the quark-model theoretical estimates. Its $D^\ast\pi\pi$
branching ratio of a few percent is a little small.
More importantly in the case of 
the orbital excitations in this mass region,  
one would expect two neighbouring resonances with the width of order several MeV
each and with the mass difference of about 30$-$50 MeV, corresponding to the $D_{2^-}$ 
and $D_{3^-}$ states. The published plots do not show any sign of a neighbouring $3^-$
resonance.

Thus we conclude that the experimental confirmation of this resonance would put forward a challenge of its
proper theoretical understanding, unless a neighbouring slightly heavier
 resonance was found.

{\bf Acknowledgements.}

Wa are grateful to Claire Bourdarios and Patrick Roudeau for attracting our attention 
to this problem and for inspiring discussions, to Damir Becirevic, Dominique Pallin 
and Alain Le Yaouanc for fruitful discussions, and to Damir Becirevic and  
Alain Le Yaouanc for presenting their results prior to publication. 
D.M. acknowledges financial support from the NATO Research Fellowships Program.

\vspace{2cm}

\begin{table}
\caption{\label{table:1}
Decay modes of possible candidate states allowed by spin-parity
conservation. Modes allowed also by the HQ symmetry, i.e. corresponding also to 
a separate conservation of the total angular momentum of the light degrees of 
freedom, are listed in bold and HQ symmetry relations for the 
corresponding amplitude squared from \protect\cite{iwprl} are given 
(without phase-space factors included).} 
\centering
\begin{tabular}{|l||l||l|l|}
\hline
            & $D'_{1^-,j=1/2}$ &  $D_{2^-,j=5/2}$   & $D_{3^-,j=5/2}$  \\
\hline
$D\pi$      & L={\bf 1}[$\frac13\beta^2$]  &$-$  & L={\bf 3}[$\frac37\alpha^2$] \\
$D^*\pi$    & L={\bf 1}[$\frac23\beta^2$]  &L=1,{\bf 3}[$\alpha^2$] & L={\bf 3}[$\frac47\alpha^2$] \\
\hline
$D^{**}\pi$: &&& \\
$D_{0^+,j=1/2}\pi$    &     $-$         & L={\bf 2}    & $-$ \\
$D_{1^+,j=1/2}\pi$    &     L={\bf 0},2 & L={\bf 2}    & L={\bf 2},4 \\
$D_{1^+,j=3/2}\pi$    &     L=0,{\bf 2} & L={\bf 2}    & L={\bf 2,4} \\
$D_{2^+,j=3/2}\pi$    &     L={\bf 2} & L=0,{\bf 2,4} & L={\bf 2,4} \\
\hline
$D(\pi\pi)_{l=0}$  & $-$   &L={\bf 2}[$\frac35\delta^2$] &  $-$                        \\
$D^*(\pi\pi)_{l=0}$& L={\bf 0},2  &L={\bf 2}[$\frac25\delta^2$] & L={\bf 2}[$\delta^2$],4 \\
\hline
$D(\pi\pi)_{l=1}$ & L={\bf 1} &L={\bf 1}[$\frac35\xi^2$],{\bf 3} & $\;\;\;\;\;\;\;$ L={\bf 3}\\
$D^*(\pi\pi)_{l=1}$& L={\bf 1},3 &L={\bf 1}[$\frac25\xi^2$],{\bf 3} &L={\bf 1}[$\xi^2$],{\bf 3},5 
\\  
\hline
\end{tabular}
\end{table}
\begin{table}
\caption{\label{table:2}
The $D^*\to D\pi$ transition.} 
\centering
\begin{tabular}{|l|l|l|l|}
\hline
             & Exp. \cite{pdg} & ${}^3P_0$ model & Other estimates \\
	     &                 &                 &(see refs in \cite{bel}) \\
\hline
$g_{D^*D\pi}$  &   $<21$  &  7$\pm$1  &  7$\div$21  \\
$\Gamma(D^*\to D\pi)$  & $<89$ KeV  & 7$\div$10 KeV & 7$\div$90 KeV  \\
\hline
\end{tabular}
\end{table}
\begin{table}
\caption{\label{table:3}
Decay rate of the $D_{2^+}\to (D,D^*)\pi$ transition.} 
\centering
\begin{tabular}{|l|l|l|}
\hline
             & Exp. \cite{pdg}& ${}^3P_0$ model \\
\hline
$\Gamma(D\pi)/\Gamma(D^*\pi)$  & 2.3$\pm$0.9    & 2.6 \\
$\Gamma_{tot}$                 & 23$\pm$5 MeV  & 11$-$22 MeV \\
\hline
\end{tabular}
\end{table}
\begin{table}
\caption{\label{table:4}
Decay rates of the $D'_{1^-}\to(D,D^*)\pi$ transition in the ${}^3P_0$ model.} 
\centering
\begin{tabular}{|l|l|}
\hline
$D'(1^-)\to D\pi$     &  150$-$220 MeV \\          
$D'(1^-)\to D^*\pi$   &  200$-$300 MeV \\
$D'(1^-)\to D^*\pi\pi$&  120$-$160 MeV \\
\hline
\end{tabular}
\end{table}

\begin{table}
\caption{\label{table:5}
Branching ratios of the $D_{2^-,j=5/2}$ decays in the ${}^3P_0$ model.} 
\centering
\begin{tabular}{|l|l|}
\hline
$Br(D\pi)$      &  $-$ \\          
$Br(D^*\pi)$    &  0.65$-$0.7  \\
\hline
$Br(D(\pi\pi)_{l=0})$   &  $< 0.01$  \\
$Br(D(\pi\pi)_{l=1})$   &  0.2$-$0.3 \\
\hline
$Br(D^*(\pi\pi)_{l=0})$ &  $< 0.003$ \\
$Br(D^*(\pi\pi)_{l=1})$ &  0.02$-$0.03 \\
\hline
$\Gamma_{tot}$      &  6$-$14 MeV \\
\hline
\end{tabular}
\end{table}
\begin{table}
\caption{\label{table:6}
Branching ratios of the $D_{3^-,j=5/2}$ in the ${}^3P_0$ model.} 
\centering
\begin{tabular}{|l|l|}
\hline
$Br(D\pi)$    &  0.65$-$0.74 \\	      
$Br(D^*\pi)$  &  0.23$-$0.27 \\
\hline
$Br(D(\pi\pi)_{l=0})$ &  $-$ \\
$Br(D(\pi\pi)_{l=1})$   & $<$0.001  \\
\hline
$Br(D^*(\pi\pi)_{l=0})$ & $< 0.003$ \\
$Br(D^*(\pi\pi)_{l=1})$ & 0.03$-$0.06\\
\hline
$\Gamma_{tot}$    &  8$-$22 MeV \\
\hline
\end{tabular}
\end{table}

\begin{thebibliography}{30}
\bibitem{delphi} DELPHI Collaboration, P. Abreu {\it et al.}, Phys. Lett. B {\bf 426} (1998) 231.  
\bibitem{gi} S. Godfrey and N. Isgur, 
Phys. Rev. D {\bf 32} (1985) 189.  
\bibitem{efg} D. Ebert, R. Faustov, and V. Galkin, 
Phys. Rev. D {\bf 57} (1998) 5663.
\bibitem{iw} N. Isgur and M. B. Wise, Phys. lett. B {\bf 232} (1989) 113; 
Phys. lett. B {\bf 237} (1990) 527.  
\bibitem{iwprl} N. Isgur and M. Wise, Phys. Rev. Lett. {\bf 66} (1991) 1130.
\bibitem{lopr} A. Le Yaouanc, L. Oliver, O. P\'ene, J.-C. Raynal, 
Phys. Rev. D {\bf 8} (1973) 2223; Phys. Rev. D {\bf 11} (1975) 1272;
Hadron transitions in the quark model, Gordon and Breach Science Publishers S.A., 
New-York, 1988.   
\bibitem{pdg} Particle Data Group, Phys. Rev. D {\bf 54} (1996) 1.
\bibitem{bel} V. M. Belyaev, V. M. Braun, A. Khodjamirian, R. R\"uckl, 
Phys. Rev. D {\bf 51} (1995) 6177. 
\bibitem{mel} D. Melikhov, Phys. Rev. D {\bf 56} (1997) 7089. 
\bibitem{damiralain} D. Becirevic and A. Le Yaouanc, paper in preparation. 
\bibitem{lwi} M.-L. Lu, M. B. Wise, N. Isgur, Phys. Rev. D {\bf 45} (1992) 1553.  
\bibitem{falk} A. Falk, in {\it Non-Perturbative Particle Theory and Experimental
Tests}, Proceedings of 20th
Johns Hopkins Workshop on Current Problems in Particle Theory, p. 79, Heidelberg, 
Germany, 1996, Edited by M. Jamin, O. Nachtmann, G. Domokos, S.
Kovesi-Domokos. World Scientific, 1997 (e-Print Archive: hep-ph/9609380).  
\bibitem{kn} A. Kaidalov and A. Nogteva, 
Sov. J. Nucl. Phys. {\bf 47} (1988) 321.         
\bibitem{exp} P. Krieger, Talk given at the ICHEP XXIX, July 1998, Vancouver, Canada; 
I. Shipsey (for the CLEO Collaboration), Talk given at the ICHEP XXIX, July 1998, Vancouver, 
Canada. 
\end{thebibliography}
\end{document}